\documentclass[useAMS,usenatbib]{mn2e}
\usepackage{graphicx,longtable,color}
\usepackage{mathrsfs}
\usepackage{epsfig}
\usepackage{natbib}
\usepackage{color}
\usepackage{float}
\usepackage{amsmath}
\usepackage{times}
\usepackage{upgreek}
\usepackage[varg]{txfonts}
\usepackage{enumerate}
\usepackage{verbatim}
\usepackage{booktabs}
\usepackage{multirow}
\usepackage{amssymb}

\title[]{Primordial alignment of elliptical galaxies in intermediate redshift clusters}
\author[Yu Rong, Shuang-Nan Zhang, Jin-Yuan Liao]{Yu Rong$^{1,2}$\thanks{E-mail: rongyu@ihep.ac.cn}, Shuang-Nan Zhang$^{1,3}$, Jin-Yuan Liao$^{1}$\\
$^{1}$Key Laboratory of Particle Astrophysics, Institute of High Energy Physics, Chinese
Academy of Sciences, Beijing, China\\
$^{2}$University of Chinese Academy of Sciences, 19A Yuquan Road, 100049 Beijing, PR China\\
$^{3}$National Astronomical Observatories,
Chinese Academy Of Sciences, Beijing, China}

\begin{document}
\maketitle

\begin{abstract}
We measure primordial alignments for the red galaxies in the sample of eight massive galaxy clusters in the southern sky from the CLASH-VLT Large Programme, at a median redshift of 0.375. We find primordial alignment with about $3\sigma$ significance in the four dynamically young clusters, but null detection of primordial alignment in the four highly relaxed clusters. The observed primordial alignment is not dominated by any single one of the four dynamically young clusters, and is primarily due to a population of bright galaxies ($M_r<-20.5\ \rm{m}$) residing in the region $300\--810$~kpc from the cluster centers. For the first time, we point out that the combination of radial alignment and halo alignment can cause fake primordial alignment. Finally, we find that the detected alignment for the dynamically young clusters is real rather than fake primordial alignment.
\end{abstract}
\section{Introduction}

Primordial alignment, also called direct alignment, is referred to as the alignment between the major axes of red galaxies and the axis of the central brightest cluster galaxy (BCG) or cD galaxy in a cluster of galaxies, and has been used as a probe of the dynamical state of clusters \citep{Plionis03}. It has been found that the alignment is stronger for dynamically young clusters \citep{Plionis02,Plionis03}, where galaxies still preserve the memory of their infall history along the large-scale filament structure within which the protocluster is embedded \citep{Wesson84,West94}. In highly relaxed clusters, where there has been sufficient time for the exchange of angular momentum of galaxies in multiple galaxy encounters in the dense cluster environment \citep{Coutts96}, one should not expect to observe any significant primordial galaxy alignment, even if they did originally exist \citep{Plionis03}. Therefore the evolution of the alignment effect offers clues to its origin and dynamical state of a cluster, and can be possibly used to constrain galaxy formation models and their interaction with large-scale structure \citep{Hung10}. Primordial alignment is also an important contamination in weak lensing measurements \citep{Hirata04}. The ellipticity of a galaxy can be subject to physical effects that stretch it and orient it in preferential directions with respect to large-scale structure \citep{Troxel14}, which can mimic the coherent galaxy alignments of gravitational lensing.

Observational efforts to detect primordial galaxy alignments in clusters or superclusters have not reached to a clear consensus. \cite{Rood72} were the first to claim that satellite galaxies in Abell~2199 tend to point in the direction of the major axis of BCG. Subsequently, primordial galaxy alignment was found in more clusters, e.g., Abell~521 \citep{Plionis03}, Abell~1689 \citep{Hung10}, Coma \citep{Djorgovski83,Kitzbichler03}, Virgo \citep{West00}, Abell~999 and Abell~2197 \citep{Adams80,Thompson76}, and in some cluster samples (Yang et al. 2006 ($0.01<z<0.2$); Plionis et al. 1994, 2003 ($z<0.15$); West et al. 1995 ($z<0.2$); Agustsson \& Brainerd 2006 (median redshift $z=0.058$); Faltenbacher et al. 2007 ($0.01<z<0.2$)). However, some other measurements have been consistent with random orientations of satellite galaxies in clusters (e.g., Hawley \& Peebles 1975; Dekel 1985; van Kampen 1990; God{\l}owski \& Ostrowski 1999; Strazzullo et al. 2005 ($z<0.27$); Torlina et al. 2007; Trevese et al. 1992; Panko et al. 2009 ($z<0.18$); Sif\'on et al. 2015 ($0.05<z<0.55$)). On one hand, the primordial alignment is primarily found in the unrelaxed clusters (e.g., Coma, Simionescu et al. 2013, Sanders et al. 2013; Abell~1689, Kawaharada et al. 2010; Virgo, Urban et al. 2011; Abell~521, Ferrari et al. 2006, Maurogordato et al. 2000, Ferrari et al. 2003), which agrees with stronger alignment in dynamically young clusters. On the other hand, at higher redshifts, the primordial alignment is denied in many clusters \citep{Sifon15} other than Abell~521 ($z\sim 0.25$), which is unexpected since the fraction of the dynamically young clusters at high redshifts is larger \citep{Weibmann13,Mann12,Maughan08,Hashimoto07}, according to the hierarchical clustering models of structure formation.

To resolve the above conflict, in this work we reexamine the issue of isotropy in galaxy position angles in the clusters located at the moderate redshifts about 0.4, where the fraction of the dynamically young clusters containing rich substructures is significant \citep{Weibmann13,Mann12,Maughan08,Hashimoto07,Bauer05}. Therefore in these dynamically young clusters, the elliptical galaxies may not have enough time to ``forget'' their primordial orientations from infall history and may show anisotropic distributions of their position angles. We describe the data and analysis in Section 2. The results are reported in Section 3 and discussed in Section 4. In Section 5, we summarize the work. We adopt the WMAP7 cosmological parameters: $\Omega_{\rm{M}}=0.27, \Omega_{\rm{\Lambda}}=0.73$, $H_0=71\ \rm km\ s^{-1}\ Mpc^{-1}$, and $h=0.7$.
\section{DATA AND ANALYSIS}
\subsection{Cluster Sample}

The clusters of the Cluster Lensing And Supernova survey with Hubble (a VIMOS Large Programme, or CLASH-VLT) project in the southern sky \citep{Piero14,Postman12} are chosen to examine the primordial alignment of the elliptical galaxies in them. The range of redshift of the clusters is $z\sim 0.2\--0.6$, with a median redshift $z\sim 0.353$ \citep{Piero14}. In order to independently examine the primordial alignments, the duplicate clusters (A209, A383, R1347.5) in the cluster samples of CLASH-VLT and \cite{Sifon15} (no alignment was detected in their sample) are abandoned. Additionally, BCGs in MACS~J1311.0-0311 and MS2137.3-2353 are nearly spherical; therefore the two clusters are also abandoned since their major axes of the BCGs cannot be identified accurately. Finally, eight clusters are left with a median redshift $z\sim 0.375$.

Although these clusters are relaxed according to their X-ray surface brightness symmetry, some evidence of merging activity and substructures has been reported in these clusters; therefore for several clusters the dynamical state is somewhat ambiguous \citep{Postman12}. More precisely, the clusters are classified to be dynamically young clusters (Y) and relaxed clusters (R) respectively, according to the following four criteria.

$\rm\uppercase\expandafter{\romannumeral1}$) The morphological code \citep{Mann12}. The assigned morphological codes (from apparently relaxed to extremely disturbed clusters) are labelled as ``1'' (pronounced cool core, perfect alignment of X-ray peak and single cD galaxy), ``2'' (good optical/X-ray alignment, concentric contours), ``3'' (nonconcentric contours, obvious small-scale substructure), and ``4'' (poor optical/X-ray alignment, multiple peaks, no cD galaxy) \citep{Sereno12}. If the morphological code of a cluster is ``3'' or ``4'', this cluster is considered dynamically young. However only the ongoing or recent merger along an axis sufficiently misaligned with our line of sight will have a code of ``3'' and ``4'' for the morphological code \citep{Mann12}; therefore, a high value of morphological code (3 or 4) is a sufficient selecting criterion for a dynamically young cluster, but a low value of morphological code (1 or 2) may not indicate a relaxed cluster. With this criterion, MACS~J0416-24, MACS~J2129-07 are selected as dynamically young clusters.

$\rm\uppercase\expandafter{\romannumeral2}$) The merging companion mass into BCGs \citep{Burke15}. BCGs grow by accreting their companions and produce the intracluster light; if a cluster is highly relaxed, we should expect it to be highly evolved, having already undergone the majority of their galaxy merging interactions and thus showing high-BCG masses, large intracluster light fractions and low-mass merging companions into BCG \citep{Burke15}. Unfortunately, the merging companion masses of the eight clusters only distribute in a small range, as listed in Table~\ref{c_inf}; therefore the merging companion mass may not be a helpful criterion for classifying clusters in this work and thus not used here.

$\rm\uppercase\expandafter{\romannumeral3}$) The projected offset between the X-ray emission peak and optical centers \citep{Umetsu14,Mann12}. The spatial segregation of gas and galaxies in a cluster (X-ray/optical offset) constitutes powerful evidence of a recent or ongoing merger \citep{Smith05,Shan10}; conversely, excellent alignment of the BCG with the intracluster gas distribution is typical of a relaxed system \citep{Mann12}. However viewing angle and merger phase of a cluster may conspire to yield a small X-ray/optical separation even for a very disturbed system \citep{Mann12}; therefore, a high value for this diagnostic is a sufficient but not a necessary selection criterion for dynamically young clusters. In this work, we select the clusters with the offset distance larger than $10\ {\rm{kpc}}/h$ as dynamically young clusters. With this criterion, MACS~J0416-24, MACS~J2129-07, and RXJ2248-4431 are selected as dynamically young clusters.

$\rm\uppercase\expandafter{\romannumeral4}$) As long as any evidence of ongoing galaxy-scale interactions or substructures is found in a cluster, this cluster is classified as dynamically young. In MACS~J1206-08, some evidence of merger activity along the line of sight is suggested by the very high velocity dispersion \citep{Postman12}, and a significant amount of intracluster light, which is not centrally concentrated, suggests that galaxy-scale interactions are still ongoing \citep{Eichner13}. Therefore, MACS~J1206-08 is indeed a dynamically young cluster. Maughan et al. (2008) found RXJ2248.7-4431 to be slightly elliptical $\epsilon=0.2\pm0.01$ and with some level of substructure. Therefore RXJ2248.7-4431 is more likely to be a dynamically young cluster.

The values of the morphological codes, merging companion mass into BCGs, and X-ray/optical offsets, and some evidence which is useful for classifying the clusters, are listed in Table~\ref{c_inf}. The eight clusters are thus divided into two subsamples, four dynamically young clusters in sample A (the median redshift $z\sim 0.418$), and four relaxed clusters in sample B (the median redshift $z\sim 0.353$).

\begin{table} \footnotesize
\begin{tabular}{@{}lccccc@{}}
\hline
\hline
Clusters & $z$ & Code & $M_{\rm{c}}$ & Offset & Type\\
& Col. (1)        & Col. (2) & Col. (3) & Col. (4) & Col. (5)\\
\hline
M0329 & 0.451 & 1 & 0.79 & 9.8 & R$^{a}$ \\
M0416 & 0.396 & 4 & 1.5 & 82.3 & Y$^{a}$ \\
M1115 & 0.353 & 1 & 0.51 & 9.5 & R$^{a}$ \\
M1206 & 0.440 & 2 & 1.1 & 8.7 & Y$^{b}$ \\
M1931 & 0.352 & 1 & -- & 4.3 & R$^{a}$ \\
M2129 & 0.588 & 3 & 1.74 & 43.3 & Y$^{a}$ \\
R2129 & 0.234 & 2 & 1.16 & 6.3 & R$^{c}$ \\
R2248 & 0.346 & -- & 1.31 & 15.9 & Y$^{d}$ \\
\hline
\hline
\end{tabular}
\caption{Information of the eight clusters. ``$\--$'' denotes no datum. Col. (1): Redshifts from Rosati et al. (2014); Col. (2): Morphological code from Mann \& Ebeling (2012); Col. (3): Merging companion mass ($\times 10^{11}M_{\odot}$) from Burke et al. (2015); Col. (4): Projected offset (kpc/$h$) between the X-ray and optical centers \citep{Umetsu14,Mann12}; Col. (5): ``Y" for a dynamically young cluster, and ``R" for a relaxed cluster. \newline
$a$. Identified by Mann \& Ebeling (2012). \newline
$b$. MACS~J1206.2-0847 appears relaxed in X-ray emission, but galaxy-scale interactions are still ongoing despite the overall relaxed state \citep{Eichner13,Postman12}. \newline
$c$. RX~J2129.6+0005 is a relaxed system confirmed by Wen \& Han (2013). \newline
$d$. Maughan et al. (2008) found RXJ2248.7-4431 to be slightly elliptical $\epsilon=0.2\pm0.01$ and with some level of substructure.}
\label{c_inf}
\end{table}

\subsection{Data Analysis}

We retrieve HST (ACS/WFC) images (30-mas scale data) of these clusters in the F475W ($g$), F625W ($r$) and F814W ($i$) bands. The associated images are fully processed and drizzled. For each cluster, photometry for those objects within its field is carried out using the Sextractor package \citep{Bertin96}. Further details about the photometry and star/galaxy classification may be found in \cite{Leauthaud07}, and we only summarize the most relevant points here. Sextractor is run twice, once with parameters appropriate to the detection and photometry of bright galaxies, without excessively deblending their images, and then with parameters suited to faint galaxies \citep{Hung10}. The same configuration parameters as in \cite{Leauthaud07} are used. The magnitudes of the sources in the cluster are calibrated by the MAG\_ZEROPOINT parameter in the AB system. The values of MAG\_AUTO and MU\_MAX of the extracted sources in the $i$ band have been used to get rid of stars with the same manner as in \cite{Leauthaud07}. The arcs and arclets, the spurious objects, and the sources in the margins of images are all visually identified and removed. The optical magnitudes of targets are corrected for the foreground extinction from the Galaxy according to the Schlegel-Finkbeiner-Davis Galactic reddening map \citep{Schlafly11}. Finally, the derived source catalogs in the three bands are matched to obtain the true objects.

\begin{figure}
\centering
\includegraphics[scale=0.4]{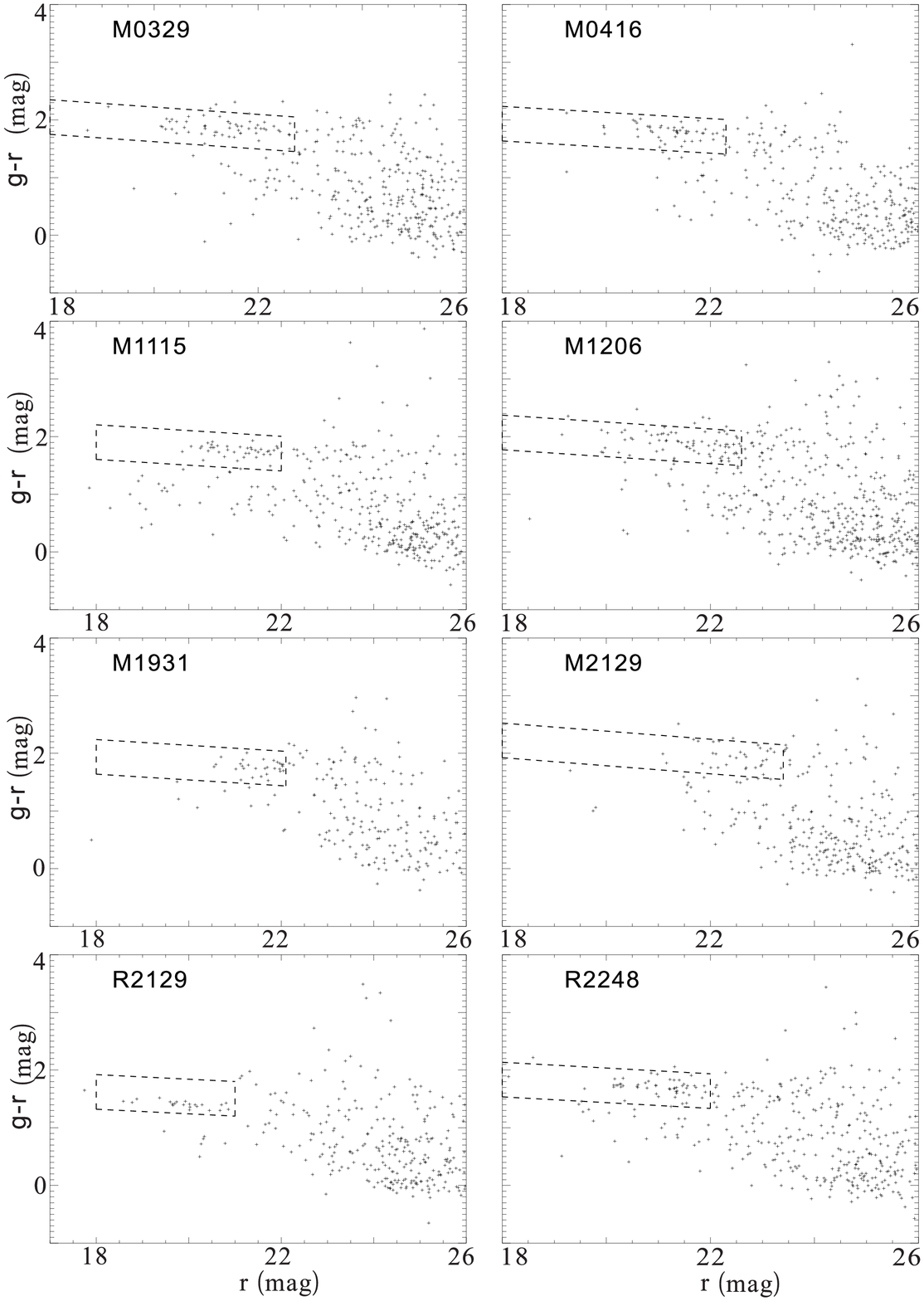}
\caption{CMDs of the eight clusters. $g-r$ color is plotted against the $r$ magnitude of the galaxies. The selected galaxies are denoted by the dashed boxes.}
\label{cmd}
\end{figure}

These objects include both the foreground and background galaxies. In order to select the red sequence \citep{Bower92}, we plot the $g-r$ versus $r$ color-magnitude diagram (CMD) for each cluster in Fig.~\ref{cmd}. The red sequence is formed by old elliptical galaxies whose spectra show similar $4000\AA$ breaks {\footnote{http://astronomy.nmsu.edu/nicole/teaching/ASTR505/lectures/lecture26/\\slide01.html}} (a characteristic of old stellar populations ubiquitous in elliptical galaxies; Pereira \& Kuhn 2005) resulting from photospheric absorptions of heavy elements \citep{Bower92}, and the $g$ and $r$ broad-band filters nicely probe the spectral region across the $4000\AA$ break at the intermediate redshift $z\sim0.4$ \citep{Presotto14}. Since the colors of elliptical galaxies are redder than spiral galaxies \citep{Baldry05}, and the Butcher-Oemler condition is satisfied, i.e., the blue galaxies are those at least 0.2 mag bluer than the cluster ridgeline \citep{Butcher84}, a linear fitting, i.e.,
\begin{equation}
(g-r)=k\cdot r+d,
\label{linear_rs}
\end{equation}
where $k$, $d$ are the slope and intercept respectively, is applied to obtain the slope of the red sequence \citep{Bower92} in the CMD, which is known as the collection of E/S0 galaxies and dwarf ``ellipticals'' \citep{Hung10}. The galaxies within the color range of $g-r\sim k\cdot r+d_{-0.3}^{+0.3}\ \rm{m}$ are selected \citep{Hung10}. The faint end of the red sequences is set as $M_{r}-5{\rm{log}}h\leq -19$ \citep{Faltenbacher07}, where $M_r$ is the absolute magnitude in the $r$-band. These selected elliptical galaxies (denoted by the dashed boxes in Fig.~\ref{cmd}) are then visually inspected in the images.

\begin{figure}
\centering
\includegraphics[scale=0.4]{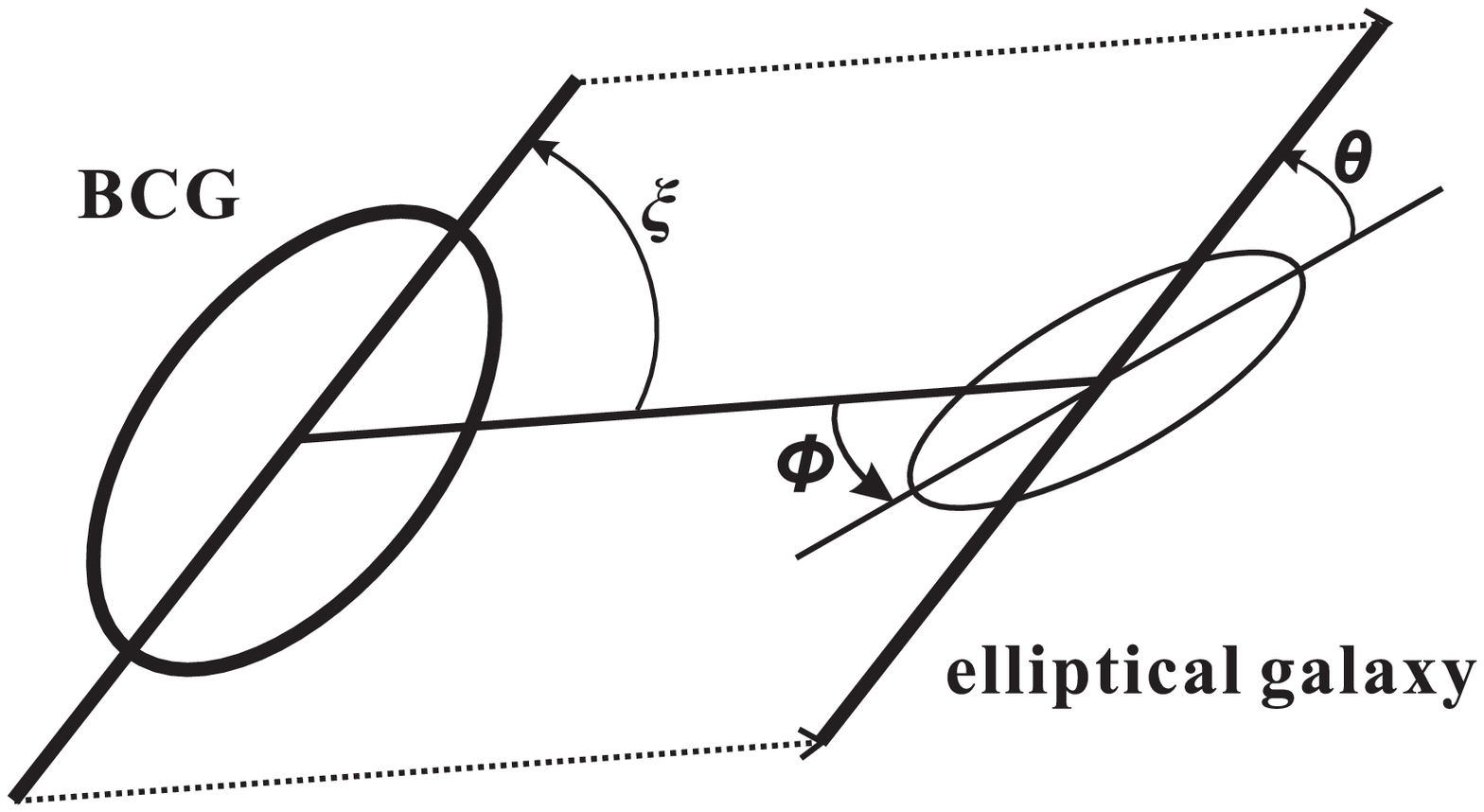}
\includegraphics[scale=0.4]{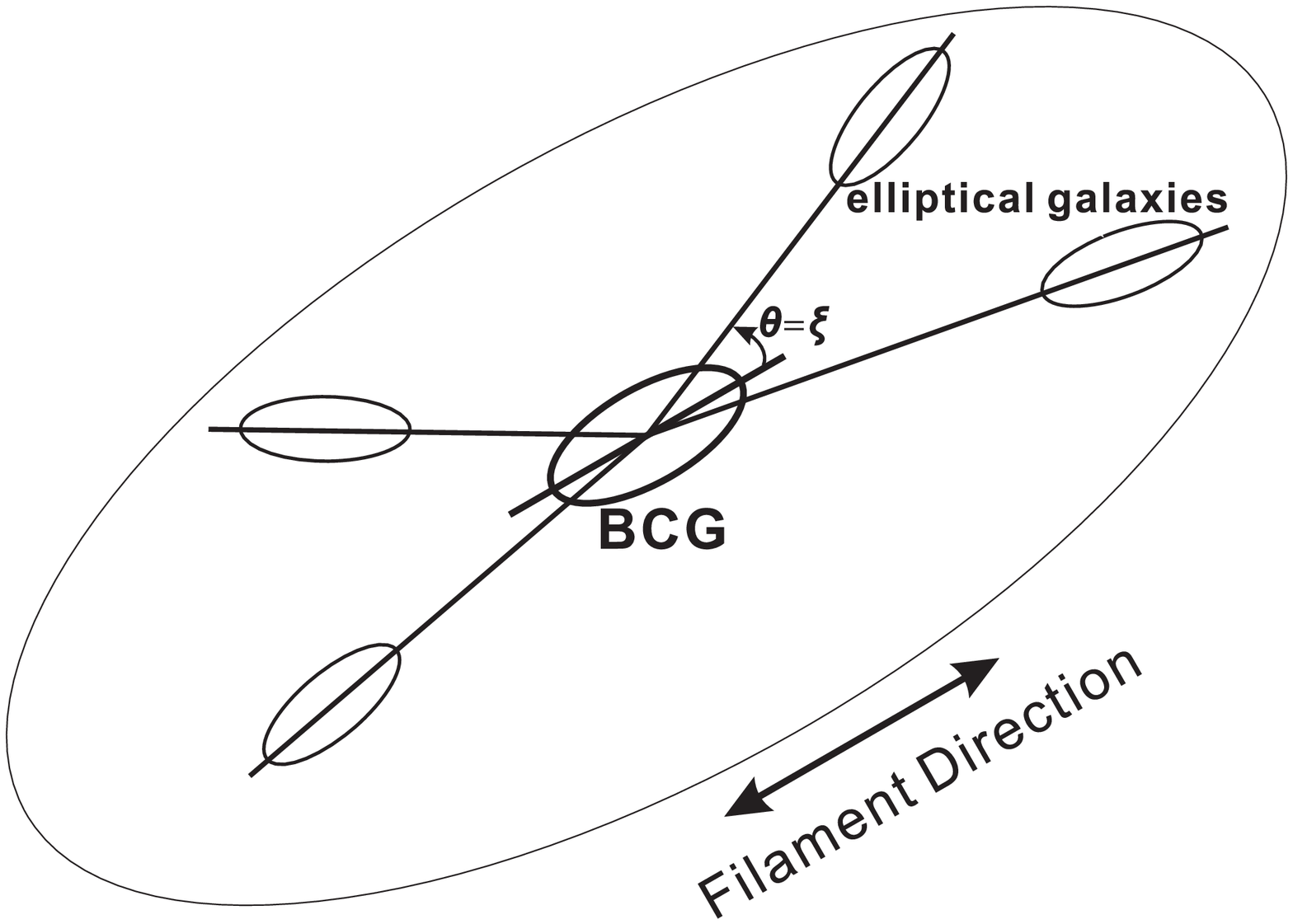}
\caption{The upper panel: illustrations of the three angles $\theta$, $\phi$, and $\xi$, which are used to test for primordial alignment, radial alignment, and halo alignment, respectively. The lower panel: illustration of fake primordial alignment caused by the combination of halo alignment and radial alignment.}
\label{ske_pa}
\end{figure}

As shown by \cite{Holden09}, it is necessary to understand the limits of our method, in order to obtain reliable measurements of the galaxy position angle. This is complicated and depends on the spatial resolution of the image, the point-spread function (PSF) and the ellipticity of galaxies \citep{Hung10}. Fortunately, our method is the same as that used by \cite{Hung10}, and the method has been tested to be robust for the targets with $\epsilon>0.2$ and $i<24\ \rm{m}$. Therefore finally only the red galaxies satisfying this criterion are selected.

In order to quantitatively measure the orientations of the detected objects and test the primordial alignment in the two samples of clusters, the position angles $\theta$ of the elliptical galaxies, as denoted in the upper panel of Fig.~\ref{ske_pa}, are calculated. First, we indentify BCG in each cluster according to their locations obtained by \cite{Donahue15}. Second, for each cluster, the orientations of major axes of the projected elliptical galaxies and BCG are characterized by the position parameter $\rm{THETA\_{SKY}}$ derived from Sextractor. Finally, the position angle of an elliptical galaxy is defined as $\theta=|{\rm{THETA\_{SKY}}}_{\rm{e}}-\rm{THETA\_{SKY}}_{\rm{BCG}}|$, where $\rm{THETA\_{SKY}}_{\rm{e}}$ and $\rm{THETA\_{SKY}}_{\rm{BCG}}$ are the position parameters of a projected elliptical galaxy and BCG, respectively. We use the average $\rm{THETA\_{SKY}}$ in the $r$ and $i$ bands to calculate $\theta$, since $r$ and $i$ bands are more likely to reflect the stellar mass distribution in an elliptical galaxy. The range of $\theta$ is from 0 to $90^{\circ}$.

If the elliptical galaxies preserve the ``infalling history'', the major axes of the projected elliptical galaxies tend to be parallel to the major axis of BCG of the host cluster and $\theta\to 0$, i.e., primordial alignment. It is worth noting that the tidal field of a cluster can cause the elliptical galaxies in a cluster radially aligned \citep{Ciotti94,Rong15}, i.e., the angles between the major axes of the elliptical galaxies and the radial directions $\phi$ (denoted in the upper panel of Fig.~\ref{ske_pa}) tend to be 0. If the galaxies in a cluster are distributed spherically symmetric, the radial alignment cannot produce a primordial alignment \citep{Rong15}. However, if the galaxies are preferentially distributed along the major axis of BCG, which is also called halo alignment \citep{Faltenbacher07,Brainerd05,Yang06,Azzaro07}, i.e., the angles between the major axis of BCG and the radial directions of the elliptical galaxies $\xi$ (denoted in the upper panel of Fig.~\ref{ske_pa}) tend to be small \citep{Faltenbacher07}, the radial alignments of the galaxies also cause the fake alignments between the major axes of the galaxies and BCG, and confuse us to identify the real primordial alignment. The fake primordial alignment caused by the combination of radial alignment and halo alignment is illustrated in the lower panel of Fig.~\ref{ske_pa}; if the galaxies are distributed along the filament (or the major axis of BCG), and meanwhile they orientate radially, then the position angles $\theta$ are equivalent to $\xi$, and also tend to be small, i.e., producing a fake primordial alignment. The possibility of fake primordial alignment should be ruled out before an alignment signal is identified to be a real primordial alignment.

\section{Primordial ALIGNMENT}

The selected elliptical galaxies are distributed within $810$~kpc from their cluster centers. Arcseconds to `kpc' are converted by \cite{Merten14}.

\begin{figure*}
\centering
\includegraphics[scale=0.7]{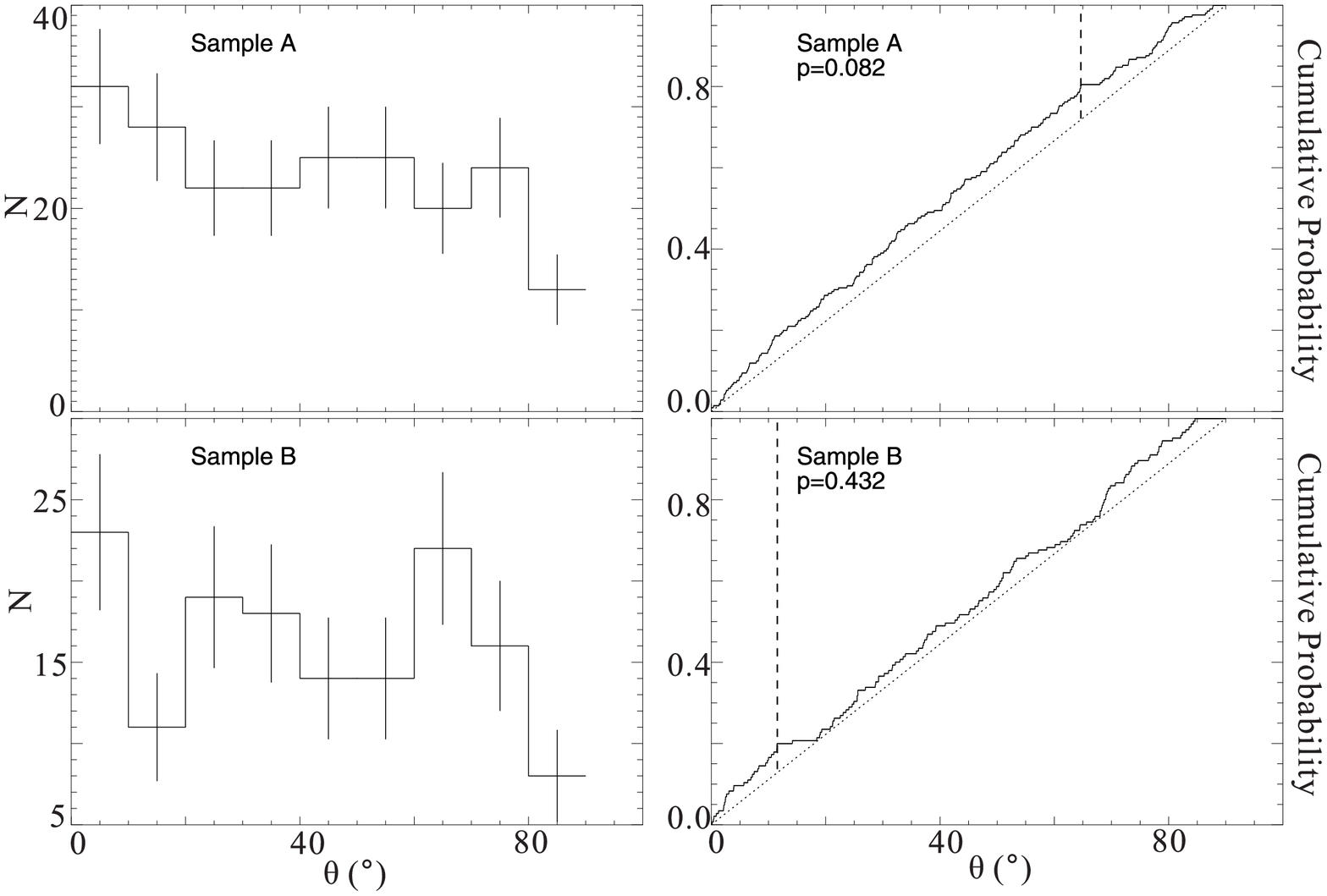}
\caption{The left panel shows PADs of the elliptical galaxies for the two samples of clusters, respectively. Hereafter the error bars assume Poisson statistics. The right panel shows CPDs (histograms) compared with the cumulative distribution function of a uniform distribution (dotted lines) for the two samples of clusters, respectively. The angles of the maximum deviations are denoted by dashed vertical lines.}
\label{pad_cd_sample}
\end{figure*}

By separating the range of $\theta$ into 9 bins, we count the number of galaxies in each bin, and plot a histogram of the position angle distribution (PAD) for each sample of clusters. The left panel in Fig.~\ref{pad_cd_sample} shows PADs of sample A (dynamically young clusters, simplified as Y), and sample B (relaxed clusters, simplified as R), respectively. In order to determine whether there is an alignment and its significance, the cumulative probability distribution of $\theta$ (CPD) compared with a uniform distribution for each sample of clusters is plotted in the right panel of Fig.~\ref{pad_cd_sample}, and the Kolmogorov-Smirnov test (K-S test) is used to detect the deviation of PAD from a uniform distribution. For each sample, the $p$ value returned by the K-S test is indicated in Fig.~\ref{pad_cd_sample}, as well as the angle of the maximum deviation from the cumulative distribution function of a uniform distribution \citep{Torlina07,Hung10}. A lower $p$ value indicates greater deviation from a uniform distribution and may imply a greater alignment if there is an excess of the number of galaxies around $\theta=0$ compared with the other bins. We find that PAD of sample A is inconsistent with a uniform distribution ($p=0.082$) and manifest a weak alignment (an excess within $\theta\simeq 0\--20^{\circ}$), while PAD for sample B is more likely to be uniform ($p=0.432$). The results indicate that there is an about $1.5\sigma$ detection of alignment in sample A (Y), while no significant deviation from isotropy in sample B (R).

\begin{figure*}
\centering
\includegraphics[scale=0.7]{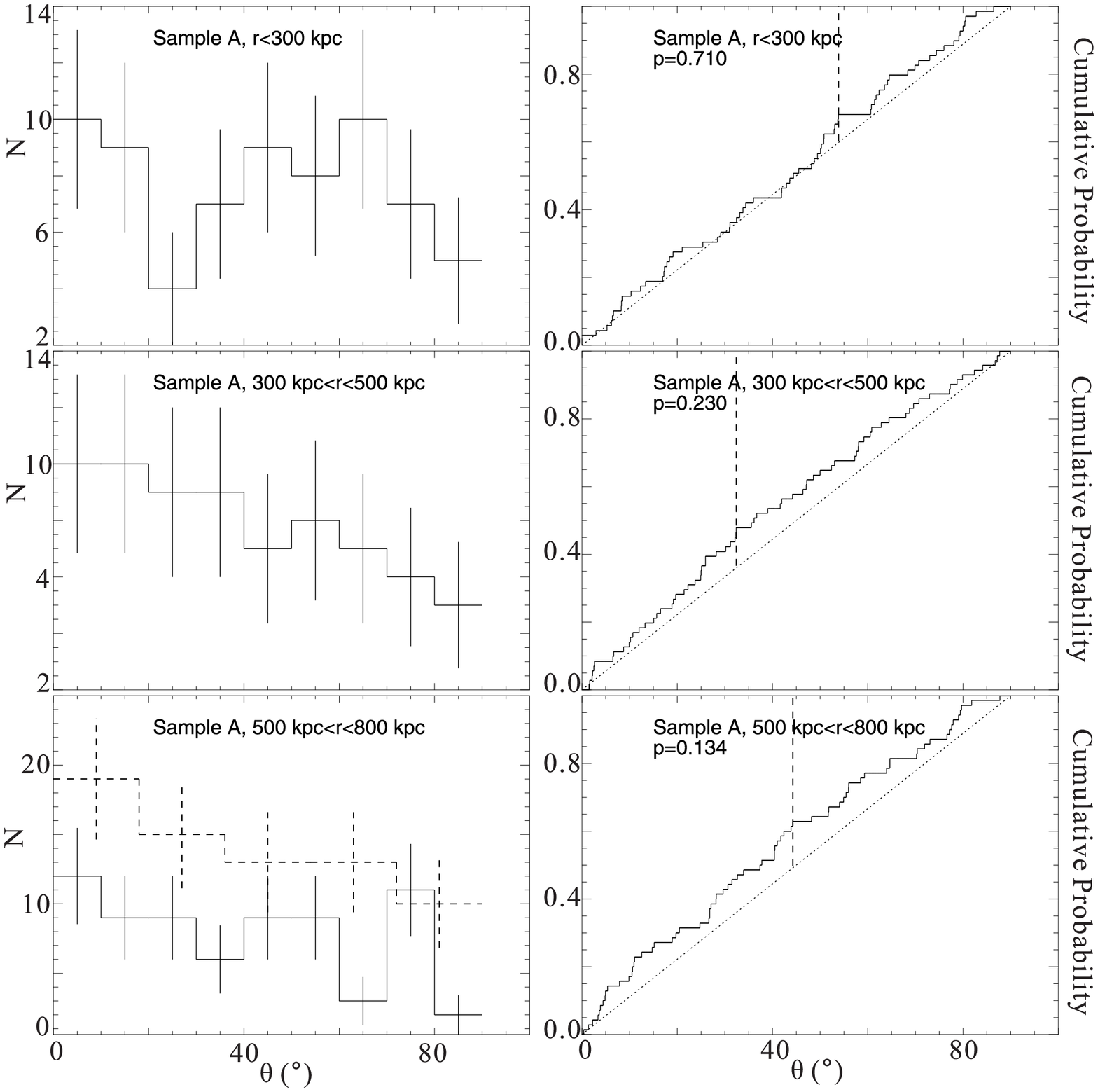}
\caption{The left panel: PADs of sample A clusters for three regions. The right panel: CPDs (histograms) compared with the cumulative distribution function of a uniform distribution (dotted lines) for sample A clusters. The angles of the maximum deviations are denoted by dashed vertical lines. From the upper panel to the lower panel, the galaxies are distributed within $0\--300$~kpc, $300\--500$~kpc, and $>500$~kpc. The dashed histogram shows PAD when we combine the nine bins into five bins.}
\label{pad_cd_Y_region}
\end{figure*}

For sample A (Y), we are also interested in where the aligned galaxies are distributed in the clusters. Therefore we plot PADs for regions within $0\--300$~kpc (projected) from the cluster centers, $300\--500$~kpc, and $>500$~kpc in the left panel of Fig.~\ref{pad_cd_Y_region}. Likewise, the results of the K-S test for the three regions are shown in the right panel of Fig.~\ref{pad_cd_Y_region}. We find that there are relatively significant alignments in the median ($p=0.230$, an excess of number of galaxies around within $\theta\simeq 0\--20^{\circ}$) and outer ($p=0.134$, small excess within $\theta\simeq 0\--20^{\circ}$ are found when we combine the nine bins into five bins) regions, but null detection of alignment in the innermost ($0\--300$~kpc, $p=0.710$) region. Therefore the alignment signal decreases toward the cluster centers.

\begin{figure*}
\centering
\includegraphics[scale=0.7]{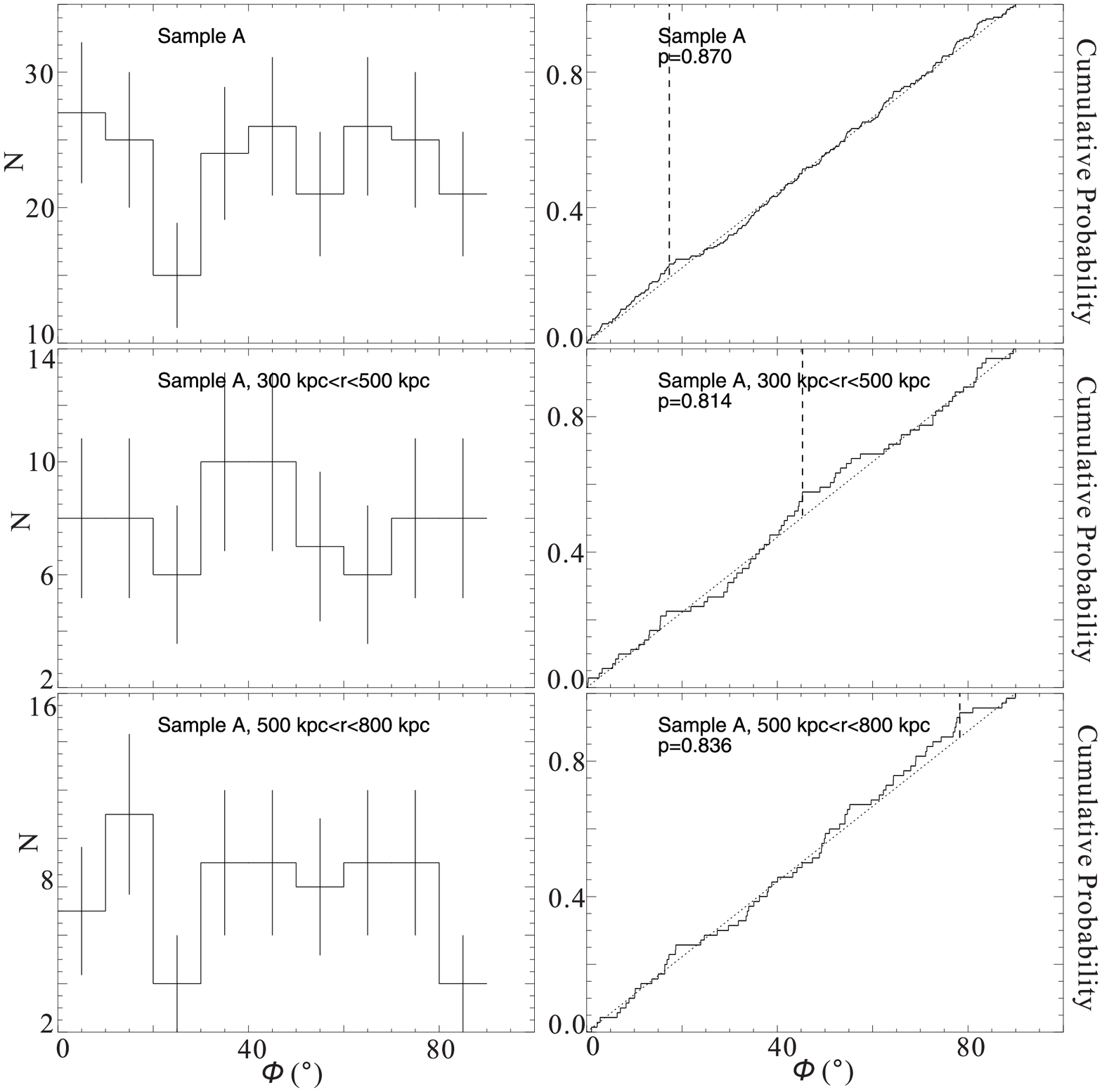}
\caption{RADs and K-S tests for the selected red galaxies in sample A distributed in the whole region (top panel), $300\--500$~kpc region (middle panel), and $>500$~kpc region, respectively.}
\label{rad_cd_Y}
\end{figure*}

In order to test whether the alignments in sample A are primordial or fake, we also examine the radial angle $\phi$ distribution (RAD), where $\phi=0$ when the major axis of a galaxy points to the cluster center, and $\phi=90^{\circ}$ when the major axis points to the tangential directions. For sample A, RAD of the galaxies in the whole region and K-S test are plotted in the top panel of Fig.~\ref{rad_cd_Y}. Analogously, RAD and K-S test for the galaxies within the regions $300\--500$~kpc and $>500$~kpc are shown in the middle and bottom panels of Fig.~\ref{rad_cd_Y}, respectively. Null detection of radial alignment is found in any of the three regions, since the probabilities that the radial angles reject the hypothesis of a uniform distribution are only 13.0\% (whole region, $p=0.870$), 18.6\% ($300\--500$~kpc region, $p=0.814$), and 16.4\% ($>500$~kpc region, $p=0.836$), respectively. Therefore, if we assume that the detected primordial alignments for sample A in the three regions are fake, then the probabilities that the position angles reject the hypothesis of a uniform distribution should be no more than $13.0\%$ for the whole region, 18.6\% for the $300\--500$~kpc region, and 16.4\% for the $>500$~kpc region, respectively; this clearly contradicts the K-S test results of 91.8\% for the whole region, 77.0\% for the $300\--500$~kpc region, and 86.6\% for the $>500$~kpc region. Therefore the detected alignments of $\theta$ cannot be caused or dominated by the combination of radial alignment and halo alignment, and thus are real primordial alignments.

\begin{figure*}
\centering
\includegraphics[scale=0.7]{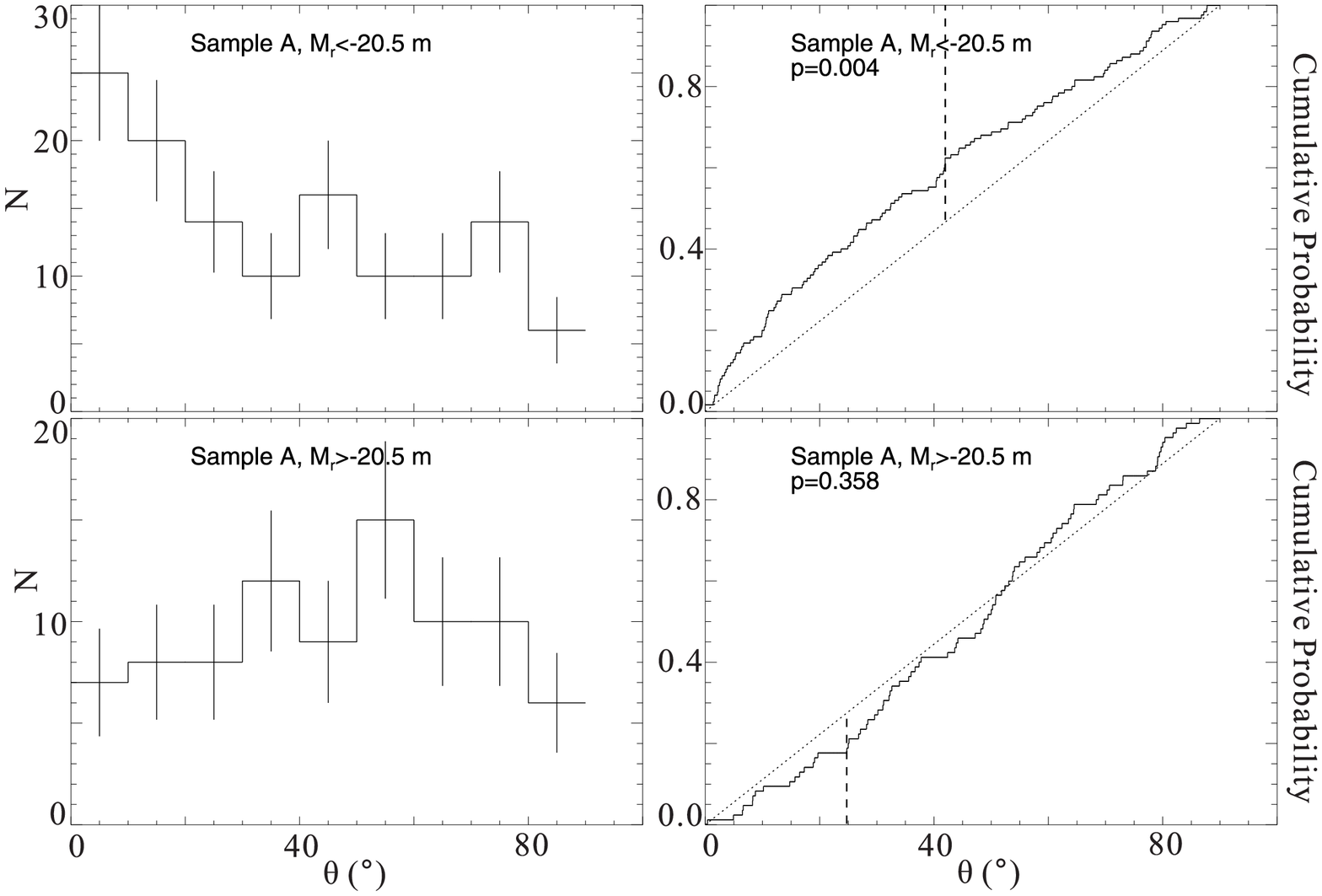}
\caption{PADs and K-S tests for the bright (upper panel) and faint (lower panel) galaxies in sample A distributed in the whole region.}
\label{pad_cd_Y_bright_faint}
\end{figure*}

We then study which galaxies are responsible for the observed primordial alignment in sample A. In Fig.~\ref{pad_cd_Y_bright_faint}, we plot the histograms of PADs and K-S tests for the bright ($M_r<-20.5\ \rm{m}$) and faint ($M_r>-20.5\ \rm{m}$) galaxies, respectively. The bright galaxies show a significant ($p=0.004$, about $3\sigma$) alignment, with a significant excess of number of galaxies around $\theta=0$. However PAD of the faint galaxies follows a uniform distribution ($p=0.358$), implying that there is no alignment of the faint red galaxies in sample A.

\begin{figure*}
\centering
\includegraphics[scale=0.7]{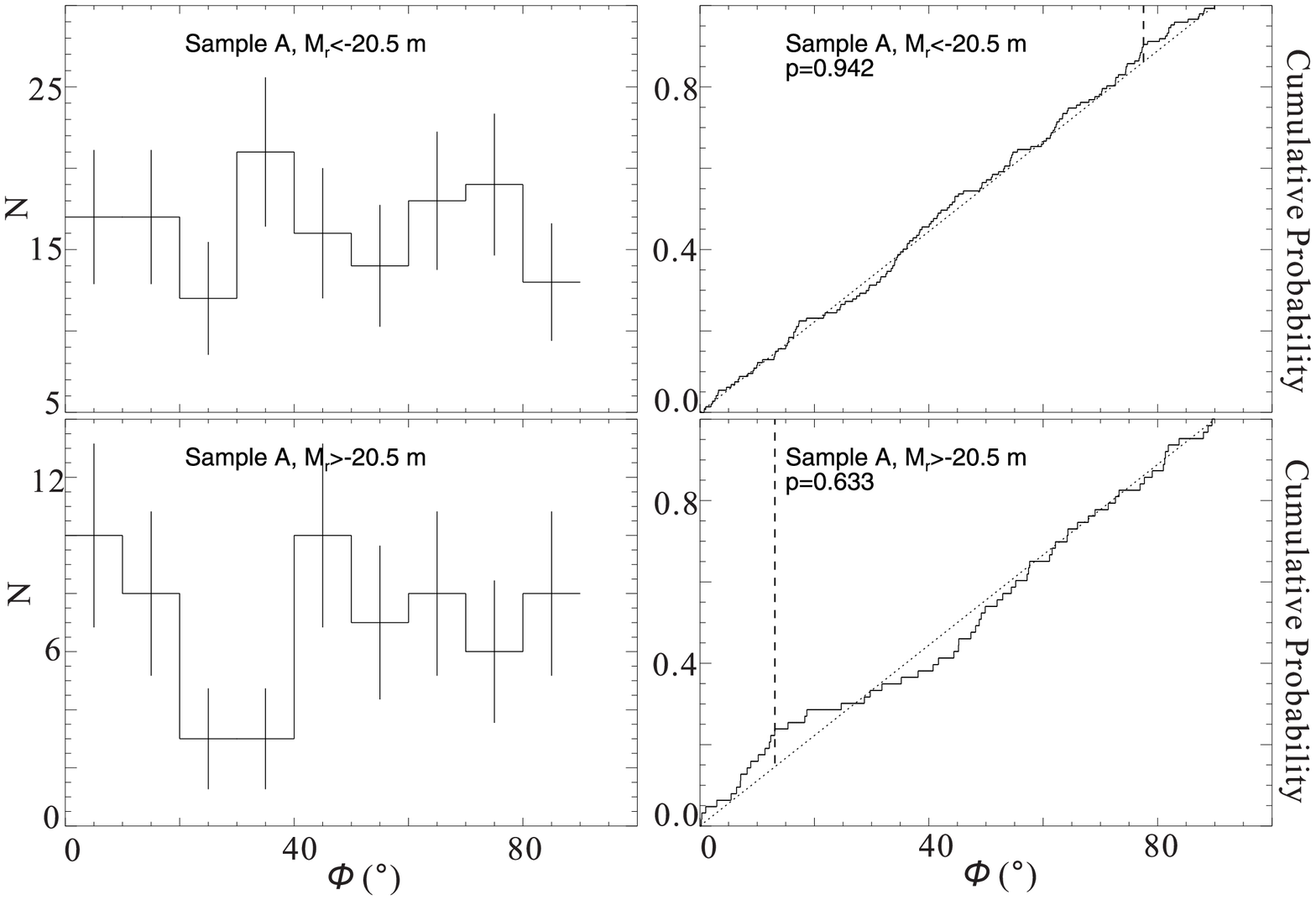}
\caption{RADs and K-S tests for the bright (upper panel) and faint (lower panel) galaxies in sample A.}
\label{rad_cd_Y_bright_faint}
\end{figure*}

Analogously, we plot RAD and the K-S test for the bright and faint galaxies of sample A in Fig.~\ref{rad_cd_Y_bright_faint}, and find no signal of radial alignment for either the bright ($p=0.942$) or faint galaxies ($p=0.633$). Finally, we conclude that the measured primordial alignment in sample A is due to a population of bright galaxies residing mainly in the region $300\--810$~kpc away from the cluster centers.

\begin{figure*}
\centering
\includegraphics[scale=0.7]{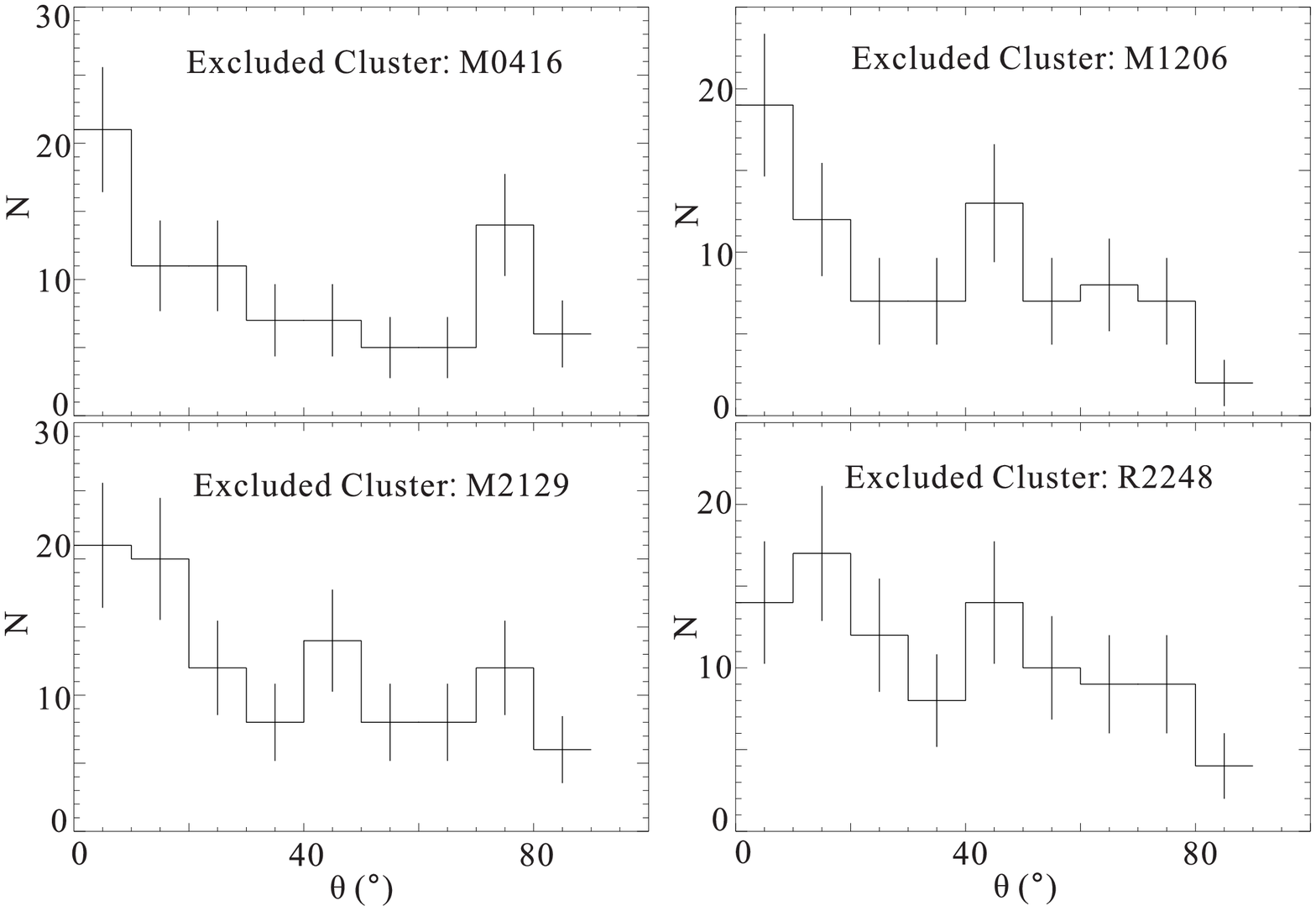}
\caption{PADs for the bright galaxies in the remanent three clusters.}
\label{pad_Y_bright_exc}
\end{figure*}

Finally, we try to test whether the detected primordial alignment signal, in particular the most significant alignment signal for bright elliptical galaxies, is dominated by any single one of the four dynamically young clusters. In each test, we exclude one cluster from sample A, and test the alignment of bright galaxies in the other three clusters. PADs of the bright elliptical galaxies in the other clusters are shown in Fig.~\ref{pad_Y_bright_exc}, and the $p$ values from K-S test are $p=0.003$ (excluding M0416), $p=0.010$ (excluding M1206), $p=0.004$ (excluding M2129), and $p=0.037$ (excluding R2248). The significance changes only between 96.3\% and 99.7\% ($2\sigma\--3\sigma$); therefore the primordial alignment signal is not dominated by any single one of the four dynamically young clusters.

\section{DISCUSSION}

Recently, \cite{Girardi15} found evidence of MACS~J1206.2-0847 having WNW-ESE elongation as the direction of the main cluster accretion, traced by passive galaxies and red strong H$\delta$ absorption galaxies. Therefore the red galaxies should be distributed along the direction of the main cluster accretion, and ``preserve'' the memory of the infall history, i.e., their major axes should be aligned with the direction of the main cluster accretion \citep{Wesson84,West94}. This agrees with the measured primordial alignment in the four dynamically young clusters including MACS~J1206.2-0847.

\cite{Plionis02} reported that dynamically young clusters are more likely aligned with their nearest neighbors and in general with the nearby clusters that belong to the same supercluster. \cite{West95} found that the substructures (in X-ray band, suggesting that they may be dynamically young) in 43 clusters of galaxies ($z<0.2$) show a tendency to be aligned with the orientations of the major axes of their parent clusters. They are strong indications that clusters develop in a hierarchical fashion by anisotropic merging along the large-scale filaments within which they are embedded. In this work, we further find that the red galaxies are more likely aligned with the major axes of BCGs in the four dynamically young clusters at the intermediate redshifts ($z\sim0.4$). Since the major axis of a BCG strongly coincides with the orientation of its parent cluster and the orientation of the nearby large-scale filamentary structure \citep{West94,Fuller99,Struble90}, therefore the primordial alignments also agree with the hierarchical models of structure formation. The observed primordial alignment conflicts with the result by \cite{Sifon15}, where null detection of primordial alignment was found in the galaxy clusters at the redshifts $0.05<z<0.55$. It is possibly due to the contamination introduced by the relaxed clusters which were not removed from their sample, since the isotropic orientations of the galaxies in those relaxed clusters dilute the possible primordial alignment in the dynamically young clusters.

In the four relaxed clusters, null detection of primordial alignment is found in this paper, since there has been sufficient time to damp the primordial alignments by violent relaxation \citep{Coutts96} or by tidal torquing of clusters \citep{Pereira08}.

The relatively significant (about $3\sigma$) primordial alignment of the bright elliptical galaxies in the sample of dynamically young clusters agrees with the observations in Virgo \citep{West00} and Coma \citep{West98}, as expected in the tidal model \citep{Pereira08}. For brighter galaxies, the change in position angle would take place slowly by tidal torquing, over multiple orbits \citep{Pereira08,Hung10}. Therefore the primordial directions of the bright galaxies from the infall history \citep{Wesson84,West94} are preserved. On the other hand, null detection of alignment is found for the faint galaxies, since the primordial position angles of the faint galaxies will be more strongly affected by the tidal fields of the clusters and would be disrupted more rapidly, {but the clusters have not been sufficiently relaxed so that the tidal fields do not have enough time to create new radial alignments \citep{Torlina07}.

The null detection of primordial alignment in the innermost region for the dynamically young clusters may be also resulted from the effect of tidal fields of the clusters. In the $0\--300$~kpc region, the original position angles of galaxies are more strongly affected by the tidal torquing \citep{Pereira08} and disrupted. In the median ($300\--500$~kpc) and outer ($500\--810$~kpc) regions, their primordial alignments are less affected. Note that the so-called ``outer'' region $500\--810$~kpc in this work is still an inner region of the CLASH clusters, compared with the virial radii of the clusters $r_{\rm{vir}}\simeq (1.5\--2)\ {\rm{Mpc}}\ h^{-1}$ \citep{Merten14,Umetsu14}. Therefore the number of the bright elliptical galaxies, which perhaps are the major sources of alignments (because the alignment signal of the bright red galaxies is the most significant), does not evidently decrease from the $300\--500$~kpc region to $500\--810$~kpc region. In the four dynamically young clusters, there are 42 bright ($M_r<-20.5 \rm{m}$) and 29 faint ($M_r>-20.5 \rm{m}$) elliptical galaxies in the $300\--500$~kpc region, and 42 bright and 28 faint elliptical galaxies in the $500\--810$~kpc region, respectively.

For the first time, we point out that the combination of radial alignment and halo alignment can cause fake primordial alignment. Therefore previous work of reporting the detections of direct alignments should be carefully reexamined to distinguish the primordial alignment from the fake alignment, since the former is resulted from the infalling history of galaxies whereas the latter also depends on the tidal torsion by the tidal fields of clusters. Particularly the fake alignment should be checked in the clusters where both the direct alignment and halo alignment are detected or in the elongated clusters, for instance, Abell~2199 \citep{Rood72}, Abell~521 \citep{Plionis03}, and Coma \citep{Schipper78,Djorgovski83}.

The detected signals of primordial alignment are not very strong (about $3\sigma$) in this work. However, the signals of alignments are so difficult to be identified that their significance is often not high; for instance, the significance is about 95.4\% in \cite{Hung10}, lower than 93.2\% in \cite{Adams80}, about 95\% in \cite{Strom78}, lower than 93\% in \cite{West00}, and about 99.4\% in \cite{Agustsson06}, since the tidal fields of clusters would disrupt the originally significant alignments of the galaxies.

\cite{Plionis03} pointed out that although telescope tracking problems can introduce artificial galaxy-galaxy alignments, they cannot create a fake alignment between galaxies and their parent cluster orientation (i.e., the major axis of BCG or cD galaxy), and thus do not affect the measurement of primordial alignment.
\section{Summary}

We examined the issue of isotropy in the position angles of elliptical galaxies in the eight clusters in the southern sky from CLASH-VLT Large Programme, at a median redshift of 0.375. The eight clusters are divided into two samples, dynamically young clusters and relaxed clusters, and the distributions and cumulative probability distributions of position angles for the two samples are shown. The K-S test is used to detect whether there is a significant deviation from a uniform distribution for the position angles of galaxies in the two samples. For a signal that significantly deviates from a uniform distribution, we study whether there is an excess of number of galaxies around $\theta=0$ to determine whether the signal manifests an alignment. For the dynamically young clusters, we also studied the significance of primordial alignments in the innermost region (0\--300~kpc), the median region (300\--500~kpc), and the outer region ($>500$~kpc), and for the bright ($M_r<-20.5\ \rm{m}$) and faint ($M_r>-20.5\ \rm{m}$) galaxies, respectively. For the first time, we pointed out that the combination of radial alignment and halo alignment can cause a fake primordial alignment. Therefore we also tested whether there are radial alignments for the dynamically young clusters. Finally, in order to test whether the primordial alignment is dominated by any single one of the four dynamically young clusters, we exclude one cluster every time, and test the alignment signals for the other three clusters. Our main results are as follows:

1. The elliptical galaxies in the four dynamically young clusters are more likely aligned with the major axes of BCGs ($p=0.082$), and null detection of primordial alignment is found in the four relaxed clusters ($p=0.432$).

2. For the dynamically young clusters, we found that there are relatively significant alignments in the 300\--500~kpc ($p=0.230$) and 500\--810~kpc ($p=0.134$) regions, but null detection of alignment in the innermost (0\--300~kpc, $p=0.710$) region.

3. The detected alignments for the four dynamically young clusters (in the whole, 300\--500~kpc and 500\--810~kpc regions) are primordial rather than fake alignments, since the radial angles in the three regions are more likely to follow a uniform distribution ($p=0.870$ for the whole region, $p=0.814$ for the 300\--500~kpc region, and $p=0.836$ for the 500\--810~kpc region).

4. The most significant ($p=0.004$) primordial alignment is found for the bright galaxies, whereas null detection of primordial alignment is found for the faint galaxies ($p=0.358$).

5. The detected primordial alignment exists in all the four dynamically young clusters ,i.e., not dominated by any single cluster.

Finally, we conclude that the measured primordial alignment, which exists in every cluster of the four dynamically young clusters, is due to a population of bright elliptical galaxies residing mainly in the region $300\--810$~kpc away from the cluster centers.

\section*{Acknowledgments}
We much thank the referee for his/her constructive and helpful suggestions. We thank Shu-Xu Yi and Yuan Liu for their helpful discussions on K-S test, and Yuan Liu for his help on foreground extinction. SNZ acknowledges partial funding support by 973 Program of China under grant 2014CB845802, by the National Natural Science Foundation of China under grant Nos. 11133002 and 11373036, by the Qianren start-up grant 292012312D1117210, and by the Strategic Priority Research Program ``The Emergence of Cosmological Structures'' of the Chinese Academy of Sciences under grant No. XDB09000000.

\bibliographystyle{mn2e}




\end{document}